\documentclass[aps,pra]{revtex4}
\usepackage{graphicx}

\begin{document}

\title{Teleportation of arbitrary $n$-qudit state with multipartite entanglement}
\author{Zhan-jun Zhang$^{a, b}$ \\
{\normalsize $^a$ Department of Physics and Center for Quantum
Information Science,} \\{\normalsize National Cheng Kung University,
Tainan 70101, Taiwan} \\{\normalsize $^b$ School of Physics \&
Material Science, Anhui University, Hefei 230039, China} }

\maketitle

\begin{minipage}{420pt}

We propose a protocol ${\cal D}_n$ for faithfully teleporting an
arbitrary $n$-qudit state with the tensor product state (TPS) of $n$
generalized Bell states (GBSs) as the quantum channel. We also put
forward explicit protocol ${\cal D}'_n$ and ${\cal D}''_n$ for
faithfully teleporting an arbitrary $n$-qudit state with two classes
of $2n$-qudit GESs as the quantum channel, where the GESs are a kind
of genuine entangled states we construct and can not be reducible to
the TPS of $n$ GBSs. \\
\end{minipage}\\

\noindent {\bf I. Introduction}\\

No-cloning theorem forbids a perfect copy of an arbitrary unknown
quantum state. How to interchange different resources has ever been
a question in quantum computation and quantum information. In 1993,
Bennett et al.[1] first presented a quantum teleportation scheme
${\cal T}_0$. In the scheme, an arbitrary unknown quantum state in
Alice's qubit can be teleported to a distant qubit $B$ with the aid
of Einstein-Podolsky-Rosen (EPR) pair. Suppose Alice has a qubit $x$
in an arbitrary unknown normalized state
\begin{eqnarray}
|\Lambda\rangle_x=\alpha|0\rangle_x+\beta|1\rangle_x,
\end{eqnarray}
where $\alpha$ and $\beta$ are complex. Alice and a remote Bob share
an EPR pair $(a, b)$, say, in the state
\begin{eqnarray}
|\Psi_0\rangle_{ab}=\frac{1}{\sqrt{2}}\sum\limits_{j=0}^1(|j\rangle|j\rangle)_{ab}.
\end{eqnarray}
This teleportaion between Alice and Bob can be seen intuitively from
the following equation,
\begin{eqnarray}
|\Lambda\rangle_x|\Psi_0\rangle_{ab}&=&\sum\limits_{i=0}^3|\Psi_i\rangle_{ax}\sigma_b^{(i)}
|\Lambda\rangle_b,
\end{eqnarray}
where $|\Psi_i\rangle_{ab}=\sigma^{(i)}_b|\Psi_0\rangle_{ab}$,
$\sigma^{(0)}=|1\rangle\langle1|+|0\rangle\langle 0|$,
$\sigma^{(1)}=|0\rangle\langle1|+|1\rangle\langle 0|$,
$\sigma^{(2)}=|0\rangle\langle1|-|1\rangle\langle 0|$ and
$\sigma^{(3)}=|1\rangle\langle1|-|0\rangle\langle 0|$. Bennett et
al's work showed in essence the interchangeability of different
quantum resources[2].

The teleportation of multi-qubit teleportation has been studied by
Lee et al[3] and Yang et al[4]. Suppose that the arbitrary $n(n\ge
2)$-qubit state Alice wants to teleport to Bob is written as
\begin{eqnarray}
|\Lambda\rangle_{x_1x_2\dots x_n}= \sum\limits_{m_N=0}^1\dots
\sum\limits_{m_2=0}^1\sum\limits_{m_1=0}^1C_{m_1m_2\dots
m_N}|m_1\rangle_{x_1}|m_2\rangle_{x_2}\dots|m_n\rangle_{x_n},
\end{eqnarray}
where $C$'s are complex coefficients and
$|\Lambda\rangle_{x_1x_2\dots x_n}$ is assumed to be normalized.
Alice and Bob share in advance $N$ same Bell states, say,
$|\Psi_0\rangle_{a_nb_n} \otimes \cdots \otimes
|\Psi_0\rangle_{a_2b_2} \otimes |\Psi_0\rangle_{a_1b_1}$. The $n$
qubits $a_1$, $a_2$, $\cdots$, $a_{n-1}$and $a_n$ is in Alice's
site. The $n$ qubits $b_1$, $b_2$, $\cdots$, $b_{n-1}$and $b_n$ in
Bob's site are used to "receive" the teleported state from Alice.
Hence, the initial joint state is
\begin{eqnarray}
|\Lambda\rangle_{x_1x_2\dots
x_n}\otimes|\Psi_0\rangle_{a_nb_n}\otimes\dots\otimes|\Psi_0\rangle_{a_2b_2}\otimes
|\Psi_0\rangle_{a_1b_1}.
\end{eqnarray}
It can be rewritten as[5]
\begin{eqnarray}
\frac{1}{2^n}\sum\limits_{i=1}^{2^n}|\Psi_{i_n}\rangle_{a_nx_n}|\Psi_{i_{n-1}}\rangle_{a_{n-1}x_{n-1}}
\cdots |\Psi_{i_1}\rangle_{a_1x_1} U_{i_ni_{n-1}\cdots
i_1;b_nb_{n-1}\cdots b_2b_1}|\Lambda\rangle_{b_1b_2\cdots
b_{n-1}b_n}.
\end{eqnarray}
If Alice performs $n$ Bell-state measurements on the qubit pairs
$(a_n,x_n$), \dots, $(a_1,x_1$) and publishes a $2n$-bit classical
message corresponding to her measurement outcomes on the qubit
pairs, then conditioned on Alice's information, Bob can recover the
arbitrary state $|\Lambda\rangle$ by performing at most $2n$
single-qubit operations. To our knowledge, as for as the
multipartite quantum state teleportation is concerned, only
protocols for $n$-qubit state teleportation are proposed[3-9], and
so far there does not exist any protocol for teleporting an
arbitrary $n$-qudit state though teleportation for one-qudit state
has ever been studied by Zubairy[10], Stenholm and Bardroff[11], and
Roa et al[12]. In this paper we will extend such sutdies. We will
directly consider the general case of teleporting an arbitrary
$n$-qudit state with the tensor product state (TPS) of $n$
generalized Bell states (GBSs) as the quantum channel.

On the other hand, recently Rigolin[6] has proposed a protocol for
teleporting an arbitrary two-qubit state with a four-particle
generalized Bell state as a genuine quantum teleportation channel
and a four-particle joint measurement. However, the multipartite
state in the Rigolin's protocol is just a tensor product state of
two Bell states in essence, not a genuine multipartite entangled
state[7]. As a consequence, the Rigolin¡¯s protocol[6] is equivalent
to the Yang-Guo protocol[4] for teleporting an arbitrary
multipartite state in principle. Very recently, Yeo and Chua[8] have
presented an explicit protocol for faithfully teleporting an
arbitray two-qubit state via a genuine 4-qubit entangled state they
constructed. They think it is an important consideration because the
four-qubit entangled state, in addition to two Bell states, could be
a likely candidate for the genuine four-partite analogue to a Bell
state. Soon later, Cheng, Zhu and Guo[9] presented a general form of
genuine multipartite entangled quantum channels for arbitrary
qubit-state teleportation. In this paper we will present an explicit
protocol for faithfully teleporting an arbitrary $n$-qudit state
with two classes of $2n$-qudit GESs, where GES is referred to as a
kind of genuine entangled states we construct and can not be
reducible to the TPS of $n$ GBSs.

This paper is organized as follows: In section II we will propose a
faithful teleportation protocol ${\cal D}_n$ of multipartite
$n$-qudit state with the TPS of $n$ GBSs as the quantum channel. In
section III we will present an explicit protocols ${\cal D}_n'$ and
${\cal D}_n''$ for faithfully teleporting an arbitrary multipartite
$n$-qudit state with two classes of GESs.  A brief summary is given in section IV.\\

\noindent {\bf II. Protocol ${\cal D}_n$ for teleporting arbitrary $n$-qudit state using TPS of $n$ GBSs}\\

Teleportation for one-qudit state has ever been studied by
Zubairy[10], Stenholm and Bardroff[11], and Roa et al[12]. However,
so far there does not exist any protocol concerning the
teleportation of an arbitrary $n$-qudit state. In this section we
will focus on this issue and propose a faithful teleportation
protocol ${\cal D}_n$ of multipartite $n$-qudit state with the TPS
of $n$ GBSs as the quantum channel.

Suppose Alice has $n$ qudits $\{X_1, X_2, \cdots, X_n\}$ in the
state of
\begin{eqnarray}
|\Lambda \rangle_{X_1X_2\cdots X_n}= \sum\limits_{j_1=0}^{d-1}
\sum\limits_{j_2=0}^{d-1}\cdots \sum\limits_{j_n=0}^{d-1}
C_{j_1j_2\cdots j_n} |j_1j_2\cdots j_n\rangle_{X_1X_2\cdots X_n},
\end{eqnarray}
where $C$'s are complex coefficients and
$|\Lambda\rangle_{X_1X_2\cdots X_n}$ is assumed to be normalized.
Moreover, Alice and Bob share in advance $n$ generalized Bell states
(GBSs) in the form
\begin{eqnarray}
|\Theta_{0000\cdots 00}\rangle_{A_1B_1A_2B_2\cdots
A_nB_n}=|\Phi_{00}\rangle_{A_1B_1}|\Phi_{00}\rangle_{A_2B_2}\cdots
|\Phi_{00}\rangle_{A_nB_n},
\end{eqnarray}
where
\begin{eqnarray}
|\Phi_{00}\rangle= \sum\limits_{j=0}^{d-1} |jj\rangle/\sqrt{d}.
\end{eqnarray}
Alice has the $n$ qudits $\{A_1, A_2, \cdots ,A_n\}$ while Bob the
$n$ qudits $\{B_1, B_2, \cdots, B_n\}$. Hence the state of the
$3n$-qudit system is
\begin{eqnarray}
|\Gamma\rangle_{X_1X_2\cdots X_nA_1 B_1A_2B_2\cdots
A_nB_n}=|\Lambda\rangle_{X_1X_2\cdots
X_n}|\Phi_{00}\rangle_{A_1B_1}|\Phi_{00}\rangle_{A_2B_2} \cdots
 |\Phi_{00}\rangle_{A_nB_n}.
\end{eqnarray}
Alice performs 2-qudit $\Phi$-state projective measurements on the
qudit pairs $(X_1, A_1)$, $(X_2, A_2)$, $\cdots$, $(X_2,A_2)$,
respectively. The 2-qudit $\Phi$-state set $\{|\Phi_{kl}\rangle_{AB}
=U^{(kl)}_A|\Phi_{00}\rangle_{AB}= V^{(kl)}_B|\Phi_{00}\rangle_{AB};
\  k,l \in \{0,1, \cdots, d-1\} \}$ is a complete orthonormal basis
set in $d^{2}$ dimensional Hilbert space for two qudits, where
\begin{eqnarray}
U^{(kl)} = \sum\limits_{j=0}^{d-1} e^{-2\pi i\overline
{|j-l|}k/d}|\overline {|j-l|}\rangle\langle j|/\sqrt{d},
\end{eqnarray}
\begin{eqnarray}
V^{(kl)} = \sum\limits_{j=0}^{d-1} e^{2\pi ijk/d} |\overline
{j+l}\rangle\langle j|/\sqrt{d},
\end{eqnarray}
\begin{eqnarray}
\overline {j+l}= (j+l) \ {\rm mod} \ d.
\end{eqnarray}
Incidentally, since $\Phi$-states can be transformed into each other
via the local unitary operations, the quantum channel linking Alice
and Bob can also be other TPSs such as $|\Theta_{k_1l_1k_2l_2\cdots
k_nl_n}\rangle_{A_1B_1A_2B_2\cdots A_nB_n}$
 instead of the TPS $|\Theta_{0000\cdots
00}\rangle_{A_1B_1A_2B_2\cdots A_nB_n}$, where
\begin{eqnarray}
|\Theta_{k_1l_1k_2l_2\cdots k_nl_n}\rangle_{A_1B_1A_2B_2\cdots
A_nB_n} &=& U^{(k_1l_1)}_{A_1}U^{(k_2l_2)}_{A_2}\cdots
U^{(k_nl_n)}_{A_n}|\Theta_{0000\cdots 00}\rangle_{A_1B_1A_2B_2\cdots
A_nB_n} \nonumber \\ &=& V^{(k_1l_1)}_{B_1}V^{(k_2l_2)}_{B_2}\cdots
V^{(k_nl_n)}_{B_n}|\Theta_{0000\cdots 00}\rangle_{A_1B_1A_2B_2\cdots
A_nB_n}.
\end{eqnarray}
After Alice's measurements, the system's state collapses to
\begin{eqnarray}
&& (|\Theta_{k_1l_1k_2l_2\cdots k_nl_n}\rangle_{A_1X_1A_2X_2\cdots
A_nX_n}\langle \Theta_{k_1l_1k_2l_2\cdots
k_nl_n}|)|\Gamma\rangle_{X_1X_2\cdots X_nA_1 B_1A_2B_2\cdots A_nB_n}
\nonumber \\
&=&(|\Phi_{k_1l_1}\rangle_{A_1X_1}\langle \Phi_{k_1l_1}|)
(|\Phi_{k_2l_2}\rangle_{A_2X_2}\langle \Phi_{k_2l_2}|) \cdots
(|\Phi_{k_nl_n}\rangle_{A_nX_n}\langle
\Phi_{k_nl_n}|) |\Gamma\rangle_{X_1X_2\cdots X_nA_1 B_1A_2B_2\cdots A_nB_n} \nonumber \\
&=&|\Phi_{k_1l_1}\rangle_{A_1X_1}|\Phi_{k_2l_2}\rangle_{A_2X_2}\cdots
|\Phi_{k_nl_n}\rangle_{A_nX_n} (_{A_1X_1}\langle
\Phi_{k_1l_1}|_{A_2X_2} \langle \Phi_{k_2l_2}|\cdots_{A_nX_n}\langle
\Phi_{k_nl_n}|) |\Gamma\rangle_{X_1X_2A_1 B_1A_2B_2\cdots A_nB_n} \nonumber\\
&=&|\Phi_{k_1l_1}\rangle_{A_1X_1}|\Phi_{k_2l_2}\rangle_{A_2X_2}
\cdots |\Phi_{k_nl_n}\rangle_{A_nX_n}(_{A_1X_1}\langle
\Phi_{00}|U^{(k_1l_1)\dag}_{A_1})(_{A_2X_2} \langle
\Phi_{00}|U^{(k_2l_2)\dag}_{A_2})\cdots
(_{A_nX_n}\langle \Phi_{00}|U^{(k_nl_n)\dag}_{A_n}) \nonumber \\
&& \times |\Lambda\rangle_{X_1X_2\cdots
X_n}|\Phi_{00}\rangle_{A_1B_1}|\Phi_{00}\rangle_{A_2B_2} \cdots
 |\Phi_{00}\rangle_{A_nB_n} \nonumber\\
&=&|\Phi_{k_1l_1}\rangle_{A_1X_1}|\Phi_{k_2l_2}\rangle_{A_2X_2}\cdots
|\Phi_{k_nl_n}\rangle_{A_nX_n} (_{A_1X_1}\langle \Phi_{00}|
 _{A_2X_2}\langle\Phi_{00}| \cdots _{A_nX_n}\langle \Phi_{00}|) \nonumber \\
&& \times |\Lambda\rangle_{X_1X_2\cdots X_n}V^{(k_1l_1)\dag}_{B_1}
|\Phi_{00}\rangle_{A_1B_1}V^{(k_2l_2)\dag}_{B_2}|\Phi_{00}\rangle_{A_2B_2}\cdots
V^{(k_nl_n)\dag}_{B_n}|\Phi_{00}\rangle_{A_nB_n}
\nonumber\\
&=&
\frac{1}{d^n}|\Phi_{k_1l_1}\rangle_{A_1X_1}|\Phi_{k_2l_2}\rangle_{A_2X_2}\cdots
|\Phi_{k_nl_n}\rangle_{A_nX_n}
V^{(k_1l_1)\dag}_{B_1}V^{(k_2l_2)\dag}_{B_2} \cdots
V^{(k_nl_n)\dag}_{B_n}|\Lambda \rangle_{B_1B_2\cdots B_n}\nonumber \\
&=&\frac{1}{d^n}|\Theta_{k_1l_1k_2l_2\cdots
k_nl_n}\rangle_{A_1X_1A_2X_2\cdots
A_nX_n}V^{(k_1l_1)\dag}_{B_1}V^{(k_2l_2)\dag}_{B_2} \cdots
V^{(k_nl_n)\dag}_{B_n}|\Lambda \rangle_{B_1B_2\cdots B_n}.
\end{eqnarray}
This means that if Alice gets the state
$|\Phi_{k_1l_1}\rangle_{A_1X_1}|\Phi_{k_2l_2}\rangle_{A_2X_2}\cdots
|\Phi_{k_nl_n}\rangle_{A_nX_n}$ via her measurement, then the state
of Bob's qudits $\{B_1, B_2, \cdots, B_n\}$ collapses to the state
$V^{(k_1l_1)\dag}_{B_1}V^{(k_2l_2)\dag}_{B_2} \cdots
V^{(k_nl_n)\dag}_{B_n}|\Lambda \rangle_{B_1B_2\cdots B_n}$. Further,
if Alice tells Bob her results (i.e., $(k_1l_1k_2l_2\cdots k_nl_n)$)
via public channel, then Bob can recover the state $|\Lambda\rangle$
in his qudits $\{B_1, B_2, \cdots, B_n\}$ by performing the local
unitary operations $V^{(k_1l_1)}_{B_1}$, $V^{(k_2l_2)}_{B_2}$,
$\cdots$, and $V^{(k_nl_n)}_{B_n}$, respectively. Up to now, we have
presented the protocol ${\cal D}_n$ for teleporting arbitrary
$n$-qudit state using TPS of $n$ GBSs. By the way, when the
dimensionality $d$ of the qudit state in our protocol ${\cal D}_n$
is 2, then the present protocol becomes the Yang-Guo protocol[4].
Further, if $n$ is equal to 2, then the present protocol becomes the
Lee-Min-Oh protocol in Ref.[3]. \\

\noindent  {\bf III. Protocols ${\cal D}'_n$ and ${\cal D}_n''$ using two classes of GESs}\\

In the last section we have shown a protocol  ${\cal D}_n$ for
teleporting arbitrary $n$-qudit state using TPS of $n$ GBSs. Now we
consider the teleportation of the same $n$-qudit state using another
entangled quantum channel between Alice and Bob as follows,
\begin{eqnarray}
|\Xi_{0000\cdots 00}\rangle_{A_1B_1A_2B_2\cdots A_nB_n} =
\Upsilon_{A_1A_2\cdots A_n}|\Theta_{0000\cdots
00}\rangle_{A_1B_1A_2B_2\cdots A_nB_n}.
\end{eqnarray}
where $\Upsilon_{A_1A_2\cdots A_n}$ is a global unitary operator
acting on the $n$ qudits $A_1, A_2, \cdots, A_n$ and can not be
reducible to $n$ local operators acting the $n$ qudits. The state of
the $3n$ qudits $X_1$,$X_2$,$\cdots$,$X_n$,$A_1$,$B_1$,$A_2$,$B_2$,
$\cdots$, $A_n$, $B_n$ is
\begin{eqnarray}
|\Gamma'\rangle_{X_1X_2\cdots X_nA_1 B_1A_2 B_2 \cdots A_n
B_n}=|\Lambda\rangle_{X_1X_2\cdots X_n}|\Xi_{0000\cdots
00}\rangle_{A_1B_1A_2B_2\cdots A_nB_n}.
\end{eqnarray}
The $2n$-qudit state set $\{|\Xi_{k_1l_1k_2l_2\cdots
k_nl_n}\rangle_{A_1B_1A_2B_2\cdots A_nB_n}=\Upsilon_{A_1A_2\cdots
A_n} |\Theta_{k_1l_1k_2l_2\cdots k_nl_n}\rangle_{A_1B_1A_2B_2\cdots
A_nB_n}; \\  k_x,l_x \in \{0,1, \cdots d\} \}$ is another complete
orthonormal basis set for $2n$ qudits. Different $\Xi$ states can be
transformed into each other via local unitary operations. Hence,
other $\Xi$ states can also be used as the quantum channel instead
of the state $|\Xi_{0000\cdots 00}\rangle_{A_1B_1A_2B_2\cdots
A_nB_n}$. Alice performs the $\Xi$-state projective measurement on
the qubits $X_1, X_2, \cdots, X_n, A_1, A_2, \cdots, A_n$ in her
site,
\begin{eqnarray}
&& |\Xi_{k_1l_1k_2l_2\cdots k_nl_n}\rangle_{A_1X_1A_2X_2\cdots
A_nX_n}\langle \Xi_{k_1l_1k_2l_2\cdots
k_nl_n}|\Gamma'\rangle_{X_1X_2\cdots X_nA_1 B_1A_2 B_2
\cdots A_n B_n}\nonumber \\
&&= |\Xi_{k_1l_1k_2l_2\cdots k_nl_n}\rangle_{A_1X_1A_2X_2\cdots
A_nX_n}(_{A_1X_1A_2X_2\cdots A_nX_n}\langle
\Theta_{k_1l_1k_2l_2\cdots k_nl_n}| \Upsilon_{A_1A_2\cdots A_n}^\dag
 \nonumber \\&& \times |\Lambda\rangle_{X_1X_2\cdots
X_n}\Upsilon_{A_1A_2\cdots A_n}|\Theta_{0000\cdots
00}\rangle_{A_1B_1A_2B_2\cdots A_nB_n}\nonumber \\
&&=\frac{1}{d^n} |\Xi_{k_1l_1k_2l_2\cdots
k_nl_n}\rangle_{A_1X_1A_2X_2\cdots A_nX_n}
V^{(k_1l_1)\dag}_{B_1}V^{(k_2l_2)\dag}_{B_2} \cdots
V^{(k_nl_n)\dag}_{B_n}|\Lambda\rangle_{B_1B_2 \cdots B_n}.
\end{eqnarray}
This indicates that if Alice obtains the state
$|\Xi_{k_1l_1k_2l_2\cdots k_nl_n}\rangle_{A_1X_1A_2X_2\cdots
A_nX_n}$ via her measurement, then the state of Bob's $n$ qudits
$B_1, B_2, \cdots, B_n$ collapses to
$V^{(k_1l_1)\dag}_{B_1}V^{(k_2l_2)\dag}_{B_2} \cdots
V^{(k_nl_n)\dag}_{B_n}|\Lambda\rangle_{B_1B_2 \cdots B_n}$. Further,
if Alice informs Bob of her results (i.e., $(k_1l_1k_2l_2\cdots
k_nl_n)$) via public channel, then Bob can recover the state
$|\Lambda\rangle$ in his $n$ qudits $B_1, B_2, \cdots, B_n$ by
performing the local unitary operations $V^{(k_1l_1)}_{B_1}$,
$V^{(k_2l_2)}_{B_2}$, $\cdots$, and $V^{(k_nl_n)}_{B_n}$,
respectively. Since $\Upsilon_{A_1A_2\cdots A_n}$ is a global
unitary operator and can not be reducible to $n$ local operators
acting on the $n$ qudits $A_1, A_2, \cdots, A_n$, the $\Xi$ states
can not be reducible to the $\Theta$ states. Hence, $\Xi$ states is
different from the TPSs of $n$ GBSs and is referred to as a kind of
genuine entangled states for it is also a candidates for teleporting
an arbitrary $n$-qudit state.

So far, we have presented the protocol ${\cal D}'_n$ for teleporting
arbitrary $n$-qudit state using a class of GESs. In the special case
of $n=2$, $d=2$ and $\Upsilon = \cos{\theta_{12}}|00\rangle\langle
00|+\sin{\theta_{12}}|11\rangle\langle
00|-\sin{\theta_{12}}|00\rangle\langle
11|+\cos{\theta_{12}}|11\rangle\langle 11|
-\sin{\phi_{12}}|01\rangle\langle 01|
 +\cos{\phi_{12}}|10\rangle\langle 01|
+\cos{\phi_{12}}|01\rangle\langle 10|
+\sin{\phi_{12}}|10\rangle\langle 10|$, the present protocol ${\cal
D}'_n$ is exactly the Yeo-Chua protocol in Ref.[8]. By the way, if
$\Upsilon_{A_1A_2\cdots A_n}$ can be reducible to $n$ local
operators acting on the $n$ qudits $A_1, A_2, \cdots, A_n$, then the
protocol ${\cal D}'_n$ is transformed into the protocol ${\cal
D}_n$.

Now let us present our protocol ${\cal D}_n''$ for teleporting an
arbitrary $n$-qudit state using another class of GESs as quantum
channel. The entangled quantum channel between Alice and Bob is as
follows,
\begin{eqnarray}
|\Xi'_{0000\cdots 00}\rangle_{A_1B_1A_2B_2\cdots A_nB_n} =
\Upsilon_{A_1A_2\cdots A_n}\Omega_{B_1B_2\cdots
B_n}|\Theta_{0000\cdots 00}\rangle_{A_1B_1A_2B_2\cdots A_nB_n}.
\end{eqnarray}
where $\Omega_{B_1B_2\cdots B_n}$ is a global unitary operator
acting on the $n$ qudits $B_1, B_2, \cdots, B_n$ and can not be
reducible to $n$ local operators acting the $n$ qudits.  Obviously,
this entangled quantum channel $|\Xi'_{0000\cdots
00}\rangle_{A_1B_1A_2B_2\cdots A_nB_n}$ is different from
$|\Xi_{0000\cdots 00}\rangle_{A_1B_1A_2B_2\cdots A_nB_n}$ used in
the protocol ${\cal D}_n'$. However, since the $n$ qudits $B_1, B_2,
\cdots, B_n$ are in Bob's site, before teleportation he can perform
the unitary operation $\Omega_{B_1B_2\cdots B_n}^\dag$. After his
performance, the quantum channel is transformed into the first class
of GESs. Surely, the teleportation can be realized. Hence the
protocol ${\cal D}_n''$ is only a slight variation of the protocol
${\cal D}_n'$ but using different quantum channels. One can easily
see that, the Chen-Zhu-Guo protocol is only our present protocol
${\cal D}_n''$ in the special case of $d=2$. By the way, if
$\Omega_{B_1B_2\cdots B_n}$ can be reducible to $n$ local operators
acting the $n$ qudits $B_1, B_2, \cdots, B_n$, then the protocol
${\cal D}_n''$ is transformed into the protocol ${\cal D}_n'$ with
other $\Xi$ state as quantum channel. \\

\noindent {\bf 4 Summary}\\

To summarize, in this paper we have presented a protocol ${\cal
D}_n$ for faithfully teleporting an arbitrary $n$-qudit state with
the TPS of $n$ GBSs as the quantum channel. Moreover, we have also
put forward explicit protocols ${\cal D}'_n$ and ${\cal D}''_n$ for
faithfully teleporting an arbitrary $n$-qudit state with two classes
of $2n$-qudit GESs as the quantum channel, respectively.\\

\noindent {\bf Acknowledgements}

This work is supported by the National Natural Science Foundation of
China under Grant Nos. 60677001 and 10304022, the science-technology
fund of Anhui province for outstanding youth under Grant
No.06042087, the general fund of the educational committee of Anhui
province under Grant No.2006KJ260B, and the key fund of the ministry
of education of China under Grant No.206063. \\

\noindent {\bf References}

\noindent[1] C. H. Bennett, G. Brassard, C. Crepeau,  R. Jozsa, A.
Peres, and W. K. Wotters, Phys. Rev. Lett. {\bf70}, 1895 (1993).

\noindent[2] M. A. Nielsen and I. L. Chuang, Quantum Computation and
Quantum Information (Cambridge University Press, Cambridge, 2000).

\noindent[3] J. Lee, H. Min, and S. D. Oh, Phys. Rev. A 66, 052318
(2002).

\noindent[4] C. P. Yang and G. C. Guo, Chin. Phys. Lett. 17, 162
(2000).

\noindent[5] Z. J. Zhang, Phys. Lett. A 351, 55 (2006).

\noindent[6] G. Rigolin, Phys. Rev. A 71, 032303 (2005).

\noindent[7] F. G. Deng, Phys. Rev.A 72, 036301 (2005).

\noindent[8] Y. Yeo and W. K. Chua, Phys. Rev. Lett. 94, 060502
(2006).

\noindent[9] P. X. Chen, S. Y. Zhu and G. C. Guo, Phys. Rev. A 74,
032324 (2006).

\noindent[10] M. S. Zubairy, Phys. Rev. A 58, 4368 (1998).

\noindent[11] S. Stenholm, P. J. Bardroff, Phys. Rev. A 58, 4373
(1998).

\noindent[12] L. Roa, A. Delgado, and I. Fuentes-Guridi, Phys. Rev.
A 68, 022310 (2003).

\enddocument